\documentclass[reprint,
superscriptaddress,
longbibliography,
nofootinbib,
nobibnotes,
amsmath,amssymb,
 aps,
 prx,
]{revtex4-2}
\usepackage[utf8]{inputenc}
\usepackage[english]{babel}
\usepackage{graphicx}
\usepackage[left=2cm,right=2cm,top=2cm,bottom=2cm]{geometry}
\usepackage{hyperref}
\usepackage{orcidlink}
\usepackage{amsfonts}
\usepackage[normalem]{ulem}
\usepackage{aas_macros}

\hypersetup{
    colorlinks=true,
    linkcolor=blue,
    filecolor=magenta,      
    urlcolor=cyan
}
\usepackage[acronym, nomain]{glossaries}
\glsdisablehyper
\setacronymstyle{long-short}
\newacronym{emri}{EMRI}{Extreme Mass Ratio Inspiral}
\newacronym{ml}{ML}{machine learning}
\newacronym{lstm}{LSTM}{long short-term memory}
\newacronym{rqa}{RQA}{recurrence quantification analysis}
\newacronym{rp}{RP}{recurrence plot}

\begin{document}

\title{Combining Machine Learning with Recurrence Analysis for resonance detection}
\begin{abstract}
The width of a resonance in a nearly integrable system, i.e. in a non-integrable system where chaotic motion is still not prominent, can tell us how a perturbation parameter is driving the system away from integrability. Although the tool that we are presenting here can be used is quite generic and can be used in a variety of systems, our particular interest lies in binary compact object systems known as extreme mass ratio inspirals (EMRIs). In an EMRI a lighter compact object, like a black hole or a neutron star, inspirals into a supermassive black hole due to gravitational radiation reaction. During this inspiral the lighter object crosses resonances, which are still not very well modeled. Measuring the width of resonances in EMRI models allows us to estimate the importance of each perturbation parameter able to drive the system away from resonances and decide whether its impact should be included in EMRI waveform modeling or not. To tackle this issue in our study we show first that recurrence quantifiers of orbits carry imprints of resonant behavior, regardless of the system's dimensionality. As a next step, we apply a long short-term memory machine learning architecture to automate the resonance detection procedure. Our analysis is developed on a simple standard map and gradually we extend it to more complicated systems until finally we employ it in a generic deformed Kerr spacetime known in the literature as the Johannsen-Psaltis spacetime.
\end{abstract}

\author{Ond\v{r}ej Zelenka\,\orcidlink{0000-0003-3639-1587}}
\email{ondrej.zelenka@asu.cas.cz}
\affiliation{Astronomical Institute of the Czech Academy of Sciences, Bo\v{c}n\'{i} II 1401/1a, CZ-141 00 Prague, Czech Republic}

\author{Ond\v{r}ej Kop\'{a}\v{c}ek\,\orcidlink{0000-0002-6489-4010}}
\affiliation{Astronomical Institute of the Czech Academy of Sciences, Bo\v{c}n\'{i} II 1401/1a, CZ-141 00 Prague, Czech Republic}
\affiliation{Faculty of Science, Humanities and Education, Technical University of Liberec, Studentsk\'{a} 1402/2, CZ-461\,17~Liberec, Czech Republic}

\author{Georgios Lukes-Gerakopoulos\,\orcidlink{0000-0002-6333-3094}}
\affiliation{Astronomical Institute of the Czech Academy of Sciences, Bo\v{c}n\'{i} II 1401/1a, CZ-141 00 Prague, Czech Republic}

\date{\today}

\maketitle

\section{Introduction}
Nonintegrable dynamical systems exhibit a high level of complexity in their dynamics due to the absence of sufficient conserved quantities (integrals of motion). Unlike a completely integrable system in which the number of integrals matches the degrees of freedom and the dynamics is always {\em regular} (and thus predictable), a nonintegrable system, in general, allows for chaotic dynamics, which makes long-term predictability impossible \cite{strogatz19}. Nevertheless, in the weakly nonintegrable system, deterministic chaos is typically not a prominent feature, and chaotic layers in its phase space are rather thin (i.e., the phase space is still dominated by regular dynamics). Instead, the most relevant hallmark of nonintegrability in a weakly nonintegrable system is the presence of so-called Birkhoff chains, which reflect the increasing width of resonances \cite{mukherjee23}. 

The width of a resonance generally grows with the strength of (still weak) nonintegrable perturbation parameter~\cite{Zelenka:2019nyp} and, especially for the most prominent resonances characterized by frequency ratios given by small integers (e.g., 1:2 or 2:3 resonances), it may grow to a significant size and occupy a considerable portion of the phase space. Therefore, these resonances may become an important factor affecting the overall dynamics of the system and must be taken into account in the analysis of trajectories that may approach the resonant region of associated Birkhoff islands of stability.

The investigation of resonant features in weakly nonintegrable dynamical systems is important for a wide range of disciplines, including various branches of fundamental and applied research and engineering. Our motivation comes from the field of relativistic astrophysics. In particular, we investigate a dynamical system consisting of a smaller body orbiting a much more massive rotating black hole. An astrophysical system of this type is not perfectly conservative, and the energy and the angular momentum of the orbiting body are slowly decreasing and radiated away in the form of gravitational waves; the orbit gradually shrinks and the system is actually evolving as an \gls{emri} \cite{amaro18}. Gravitational waves emitted by \glspl{emri} will be detectable by future gravitational observatories (e.g., LISA \cite{amaro23}). 

Nevertheless, the complete integrability of the \gls{emri} system is only guaranteed as far as the idealization of the orbiting point mass is concerned. This is hardly the case, and integrability will be inevitably perturbed to some degree, and Birkhoff chains will grow in the phase space. The question arises whether and how such prolonged resonances induced by nonintegrable perturbations \cite{LG:2021hgwa} (caused, e.g., by the presence of extra matter around the central object \cite{Sukova:2013MNRAS} or by finite size and rotation of the smaller body \cite{hartl03}) may affect emitted gravitational waveforms and detectability of \glspl{emri} modeled as systems with three or more non-reducible degrees of freedom. 

To this end, we first develop a robust and generally applicable technique for the detection of resonances in the phase space of a (weakly) nonintegrable dynamical system with an arbitrary number of degrees of freedom. Our method combines the benefits of recurrence analysis \cite{Marwan:2007rps} with the machine learning approach. In the current paper, we first describe the method and demonstrate its abilities on simple systems (standard and de Vogeleare maps). Then we shift to a 4D map, to show that our method works on higher dimensional systems. Finally, we employ our method on a generically deformed Kerr black hole spacetime~\cite{Johannsen2011PhRvD} as an application on the astrophysical problem of \glspl{emri}.

The rest of the article is organized as follows. Our experimental setup is provided in Sec.~\ref{sec:ExpSetUp}, in which we brief an introduction into dynamical systems, recurrence analysis, embedding, the employed machine learning architecture and its training. Sec.~\ref{sec:Results} presents our obtained results from the machine learning method we employed. Our conclusions are given in Sec.~\ref{sec:Conc}.

\section{Experimental setup}\label{sec:ExpSetUp}

\subsection{Dynamical systems}

\begin{figure}
    \centering
    \includegraphics[width=\linewidth]{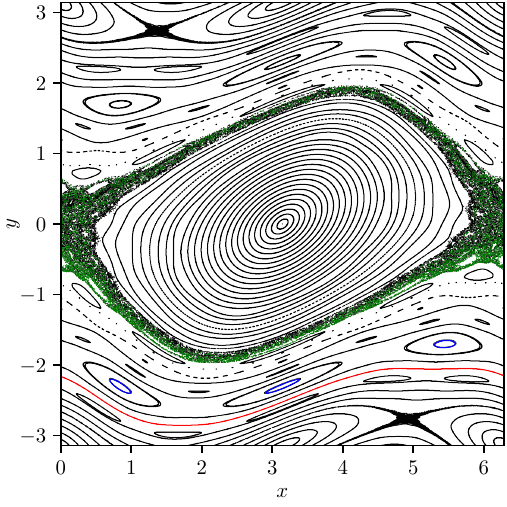}
    \caption{Phase portrait of the standard map (see Eqs.~\ref{eqs:smap}) with $K=0.8$, showing the standard hallmarks of chaotic maps: KAM curves, Birkhoff chains, and chaotic layers. The regular, resonant, and chaotic orbits which produce the \acrlongpl{rp} of Fig.~\ref{fig:example_rps} are highlighted in red, blue, and green, respectively.}
    \label{fig:example_smap}
\end{figure}

\subsubsection{Hamiltonian systems}

Many systems in physics can be written in the form of a Hamiltonian system. With $d$ degrees of freedom, it consists of
\begin{itemize}
    \item generalized coordinates $q^i,\,i=1,\hdots,d$~,
    \item canonical momenta $p_i,\,i=1,\hdots,d$~,
    \item a Hamiltonian function $H\left(q^i,\,p_i\right)$~.
\end{itemize}
The equations of motion then have the form
\begin{subequations}
\begin{alignat}{2}
    \frac{\mathrm{d}q^i}{\mathrm{d}t} &=  &&\frac{\partial H}{\partial p_i} ~,\\
    \frac{\mathrm{d}p_i}{\mathrm{d}t} &= -&&\frac{\partial H}{\partial q^i} ~.
\end{alignat}
\end{subequations}
In such a system it holds that the Hamiltonian $H$ is itself an integral of motion.

A special case is a so-called \textit{integrable} system. When $d$ linearly independent conserved quantities $F_1,\hdots,F_d$ in involution
\begin{equation}
    \frac{\mathrm{d}F_i}{\mathrm{d}t} = 0 ~,\quad \sum_{k=1}^d\left(\frac{\partial F_i}{\partial q^k}\frac{\partial F_j}{\partial p_k} - \frac{\partial F_i}{\partial p_k}\frac{\partial F_j}{\partial q^k}\right) = 0 ~.
\end{equation}
exist, there exists a set of \textit{action-angle coordinates} $\theta^i,\,\omega_i$ such that
\begin{equation}
    \frac{\mathrm{d}\theta^i}{\mathrm{d}t} = \omega_i ~,\quad \frac{\mathrm{d}\omega_i}{\mathrm{d}t} = 0 ~.
\end{equation}
The motion then takes place on a set of invariant tori which foliate the entire phase space. The actions $\omega_i$ correspond to \textit{fundamental frequencies} of the individual degrees of freedom. A torus whose fundamental frequencies are commensurable (i.e. their ratio is a rational number) consists of periodic orbits and is called \textit{resonant}.

The application of a small perturbation transforms the phase space into a \textit{near-integrable} system, which follows two elementary theorems:
\begin{itemize}
    \item the Kolmogorov-Arnold-Moser theorem \cite{Kolmogorov54,Arnold63,Moser62}: some of the non-resonant tori survive deformed,
    \item the Poincar\'{e}-Birkhoff theorem \cite{Birkhoff13}: the resonances form thinner tori which cross the surface of section multiple times before connecting back to the initial circle, surrounded by a thin chaotic zone.
\end{itemize}

\subsubsection{Dynamical maps}

The structure of a $d=2$ Hamiltonian system's phase space is commonly visualized using a \textit{Poincar\'{e} section}. A surface of section is defined in the phase space such that the Hamiltonian flow is not tangent at any point of the surface, and successive intersections of the orbit and the surface of section are recorded and form the Poincar\'{e} section. The mapping of an initial condition on the surface of section to the next intersection is called the return mapping.

In an integrable system, all successive interactions of an orbit with the surface of section lie on the intersection of the corresponding invariant torus with the surface of section, which is topologically a circle. In the case of a resonant torus, the intersections form isolated points, while a quasiperiodic torus produces gradually fills the circle densely.

The typical behavior of return mappings of near-integrable systems, such as a main island of stability containing chaotic layers and Birkhoff chains surrounded by a chaotic sea, can be observed in general dynamical maps.

This includes the return mapping of a Hamiltonian system, which is thus also a dynamical map.

\subsection{Recurrence analysis}

Recurrence analysis is based on systematic observations of a dynamical system visiting a similar state repeatedly. The \gls{rp} of a system is based on the recurrence matrix~\cite{Marwan:2007rps}
\begin{equation}
    \mathbf{R}_{i,j} = \begin{cases}
        1\,: &\left|\vec{x}_i - \vec{x}_j\right| \leq \epsilon ~,\\
        0\,: &\left|\vec{x}_i - \vec{x}_j\right| > \epsilon ~,\\
    \end{cases} ~i,\,j = 1,\dots,N~,
\end{equation}
Clearly, the main diagonal is filled with ones, as $\forall i~\left|\vec{x}_i - \vec{x}_i\right| = 0$. This is a structure which is undesirable for some of the further steps, therefore a \textit{Theiler window} is applied, which zeros out all elements on the main diagonal. Optionally, this can apply to elements up to a fixed distance from the main diagonal; however, this is a feature not used in this paper.

\begin{figure*}
    \centering
    \includegraphics{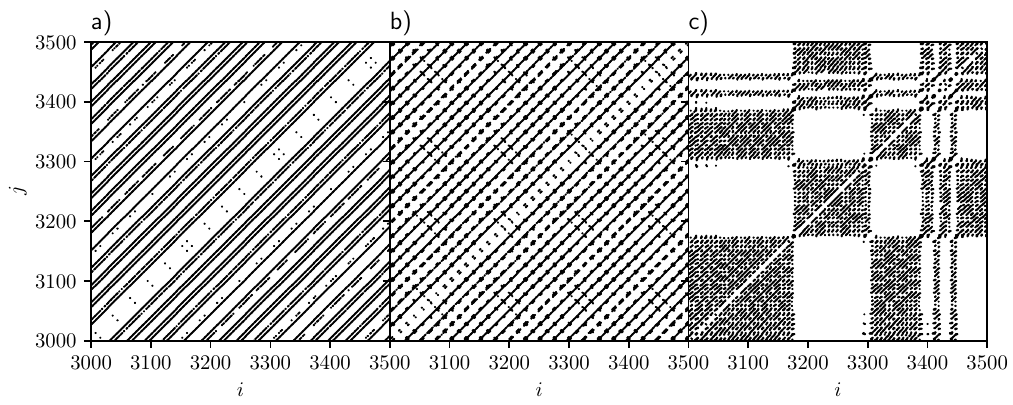}
    \caption{Examples of recurrence plots of trajectories of the standard map with $K=0.8$. Different recurrence thresholds are used in order to achieve a fixed $\mathrm{RR} = 0.05$, and all three are generated using the Euclidean metric. a) KAM trajectory starting at $\left(x_0,\, y_0\right) = \left(\pi,\, -4\pi/5\right)$, using $\epsilon = 0.16471$. b) Resonant trajectory starting at $\left(x_0,\, y_0\right) = \left(\pi,\, -37\pi/50\right)$, using $\epsilon = 0.045881$. c) Chaotic trajectory starting at $\left(x_0,\, y_0\right) = \left(\pi,\, -27\pi/50\right)$, using $\epsilon = 0.286985$.}
    \label{fig:example_rps}
\end{figure*}

A visualization of the recurrence matrix using black/white dots for the 1/0 values is called a \gls{rp}. Provided a suitable choice of parameters, it can reveal many properties of a dynamical system: a quasiperiodic orbit forms diagonal lines that are offset from each other by the time the system takes to return to an $\epsilon$ distance from a previous state as seen in the left panel of Fig.~\ref{fig:example_rps}. A chaotic orbit, on the other hand, forms structures that are generally less organized but resemble those of a quasiperiodic orbit when in a sticky zone. These manifest themselves as separated squares along the main diagonal, see the right panel of Fig.~\ref{fig:example_rps}.

Besides visual survey of \glspl{rp}, one can also perform \gls{rqa} and employ the recurrence matrix $\mathbf{R}_{i,j}$ to calculate various statistical measures related to characteristic structures appearing in \glspl{rp}. In particular, we make use of the \gls{rqa} measures based on the distributions of diagonal lines of length $l$,  $P\left(\epsilon, l\right)$, and vertical lines of length $v$, $P\left(\epsilon, v\right)$ which are defined as follows:
\begin{widetext}
\begin{subequations}
\begin{align}
    P\left(\epsilon, l\right) &= \sum_{i,j=1}^N \left(1-\mathbf{R}_{i-1,j-1}\left(\epsilon\right)\right)\left(1-\mathbf{R}_{i+l,j+l}\left(\epsilon\right)\right) \prod_{k=0}^{l-1}\mathbf{R}_{i+k,j+k}\left(\epsilon\right) ~,\\
    P\left(\epsilon, v\right) &= \sum_{i,j=1}^N \left(1-\mathbf{R}_{i,j}\left(\epsilon\right)\right)\left(1-\mathbf{R}_{i,j+v}\right) \prod_{k=0}^{v-1}\mathbf{R}_{i,j+k} ~.
\end{align}
\end{subequations}
\end{widetext}

First of all, the recurrence rate $\mathrm{RR}$ is calculated to measure the relative number of recurrences (i.e., the average density of black dots in the respective \gls{rp}):
\begin{equation}
        \mathrm{RR} = \frac{1}{N^2}\sum_{i,j=1}^N \mathbf{R}_{i,j}\left(\epsilon\right) ~.\label{eq:def_rr}
\end{equation}
The value of $\mathrm{RR}$ has an obvious dependence on threshold $\epsilon$. Higher $\epsilon$ means more recurrences and higher $\mathrm{RR}$. In some applications, it appears useful to use the fixed recurrence rate option, i.e., to first specify the value of $\mathrm{RR}$ and find the corresponding $\epsilon$, which is then used to calculate other \gls{rqa} measures and to construct the \gls{rp}.

In our analysis we employ the following \gls{rqa} indicators based on the distribution of diagonal lines:
\begin{itemize}
    \item determinism
    \begin{equation}
        \mathrm{DET} = \frac{\sum_{l=l_\mathrm{min}}^N lP\left(l\right)}{\sum_{l=1}^N lP\left(l\right)} ~,
    \end{equation}
    \item average diagonal line length (of minimal length $l_\mathrm{min}$)
    \begin{equation}
        L = \frac{\sum_{l=l_\mathrm{min}}^N lP\left(l\right)}{\sum_{l=l_\mathrm{min}}^N P\left(l\right)} ~,
    \end{equation}
    \item longest diagonal line and divergence
    \begin{equation}
        L_\mathrm{max} = \max\left(\{l_i\}_{i=1}^{N_l}\right) ~,\quad \mathrm{DIV} = \frac{1}{L_\mathrm{max}} ~,
    \end{equation}
    \item entropy of the diagonal line distribution
    \begin{equation}
        L_\mathrm{entr} = - \sum_{l=l_\mathrm{min}}^N P\left(l\right)\log P\left(l\right) ~,\label{eq:def_l_entr}
    \end{equation}
\end{itemize}
and the following \gls{rqa} measures based on the distribution of vertical lines:
\begin{itemize}
    \item laminarity
    \begin{equation}\
        \mathrm{LAM} = \frac{\sum_{v=v_\mathrm{min}}^N vP\left(v\right)}{\sum_{v=1}^N vP\left(v\right)} ~,\label{eq:def_lam}
    \end{equation}
    \item and the longest vertical line
    \begin{equation}
        V_\mathrm{max} = \mathrm{max}\left(\{v_l\}_{l=1}^{N_v}\right) ~.
    \end{equation}
\end{itemize}
The explicit dependence of the above indicators on threshold $\epsilon$ is not stressed anymore for the brevity. The values of $l_\mathrm{min}$ and $v_\mathrm{min}$ are independent parameters which specify the lengths of the shortest diagonal and vertical lines considered for the analysis.  Both values are set to 2 by default. For more details and definitions of another \gls{rqa} measures, we refer to \cite{Marwan:2007rps}.

The \gls{rqa} quantifiers encode many dynamical properties of the underlying system, such as the correlation entropy and dimension, or the maximum Lyapunov exponent~\cite{Marwan:2007rps}. Nevertheless, the relations between these dynamical invariants and \gls{rqa} measures are not straightforward due to their dependence on parameters of the recurrence analysis \cite{marwan11}. In particular, \gls{rqa} quantifiers do not directly recognize whether the trajectory is resonant (i.e., belonging to the particular Birkhoff chain) or nonresonant (i.e., represented by a KAM curve in the two-dimensional phase portrait).

Recurrence analysis proved very useful in analyses of strongly nonintegrable systems \cite{glg18,kopacek10} in which the Birkhoff chains are typically surrounded by prominent chaotic regions ("islands of stability in the chaotic sea"). In such a case, the detection and localization of the resonant region in the phase space is based on the distinction between chaotic and resonant quasiperiodic orbits. For this task, the recurrence analysis appeared very effective, allowing us to distinguish between both dynamic regimes on a shorter time scale compared to standard tools like Lyapunov characteristic exponents. 

Nevertheless, in a weakly nonintegrable system, the problem becomes more subtle as the Birkhoff chains are typically surrounded by KAM tori of (nonresonant) quasiperiodic orbits (very thin chaotic layers are not relevant in this context). In order to localize the resonance, we, therefore, need to unambiguously distinguish between the regular quasiperiodic orbits belonging to the resonance and nearby KAM tori. In a system of two degrees of freedom, the method of choice for this specific task would be a rotation number, which is directly related to the characteristic frequency ratios of the orbits. However, for the system of three or more degrees of freedom, the rotation number (computed from the Poincar\'{e} surface of section) has not been defined yet, and more general methods need to be applied. We propose employing recurrence analysis, which is applicable regardless of the number of degrees of freedom. On the other hand, the relations of \gls{rqa} measures to frequencies of the orbit are not direct, and they are further obscured by dependence on the parameters of the recurrence analysis (in particular, the threshold $\epsilon$).

\begin{figure*}
    \centering
    \includegraphics[width=\linewidth]{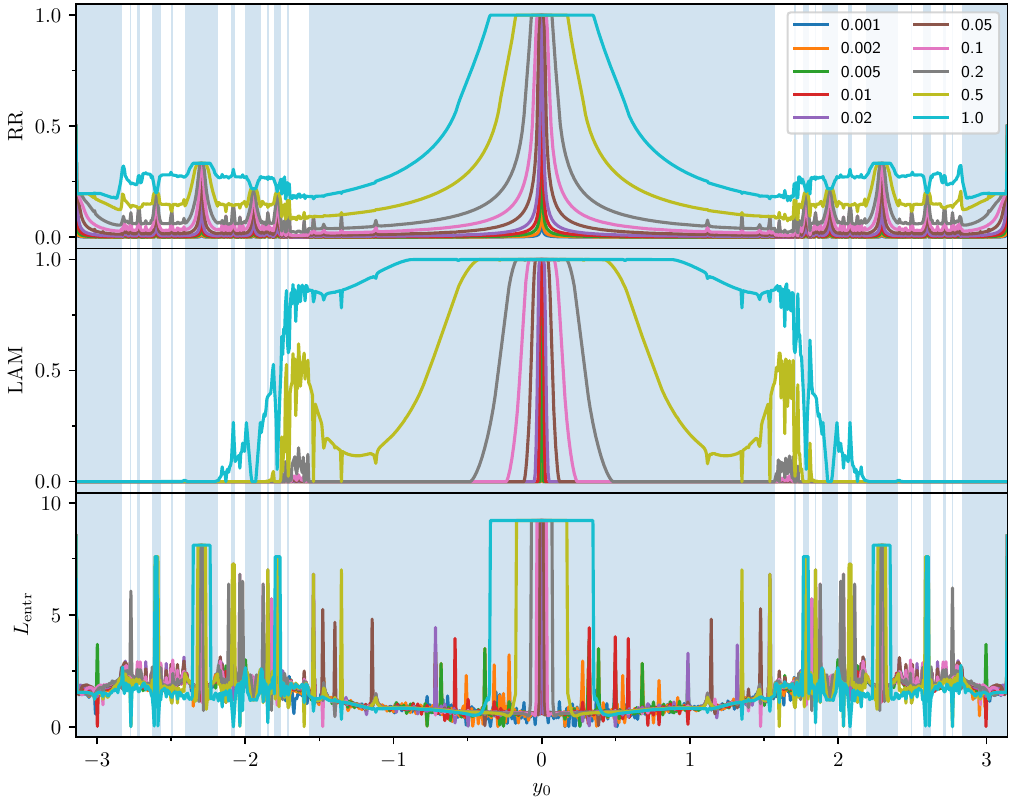}
    \caption{Example of \gls{rqa} curves of the standard map at $K=0.8$ as a function of the initial condition taken along the $y$-axis of Fig.~\ref{fig:example_smap} with $x_0=0$. Each of the three panel corresponds to one of the recurrence quantifiers: the recurrence rate, the laminarity, and the entropy of the diagonal lines distribution, in Eqs.~\eqref{eq:def_rr}, \eqref{eq:def_lam}, \eqref{eq:def_l_entr}, respectively. The color-coding corresponds to different recurrence thresholds given in the legend. The pale blue areas in the background show the locations of resonances as marked during the training data labeling procedure.}
    \label{fig:example_rqa_curves}
\end{figure*}

However, as one can observe in Fig.~\ref{fig:example_rqa_curves} which shows the dependence of several \gls{rqa} measures on initial conditions of trajectories for several values of $\epsilon$, the resonances clearly do affect the values of recurrence quantifiers, and their behaviour thus in principle allows to distinguish between resonant and nonresonant orbits and measure the width of the resonance. Nevertheless, unlike rotation numbers which remain constant within resonances and allow to detect resonances easily as plateaus in respective rotation curves, the interpretation of \gls{rqa} curves is more complicated. In particular, the values of \gls{rqa} parameters are not constant within the resonance, and their behavior depends critically on the threshold $\epsilon$ whose optimal value is not known a priori.

For example, the recurrence rate values in the top panel of Fig.~\ref{fig:example_rqa_curves} clearly respond to resonances; however, while they show peaks in most cases, the values in the resonance at approximately $y_0\in \left[-2,\,-1.9\right]$ follow this behavior for most thresholds but the opposite is true for the curve with recurrence threshold $\epsilon=1$. Thus, one should always take into account ensembles of many \gls{rqa} curves with different values of $\epsilon$ covering a sufficiently wide range of recurrence rates ($\approx 1 - 20\%$).  Moreover, in order to fully exploit the potential of recurrence analysis, we want to gather and combine information from the behavior of several \gls{rqa} measures. In order to achieve this, we propose to employ \gls{ml} methods.

\subsection{Embedding}

Recurrence analysis as described above relies on our knowledge of all components of the state vector. When applied to data where this is not true, such as observations of radiation from a complex source, spurious correlations pollute the recurrence matrix and reduce the faithfulness of the analysis.

To mitigate this issue, phase space reconstruction is necessary. The most common way to do this is to apply time delay embedding~\cite{Marwan:2007rps, Takens:1981emb}, which replaces the unknown components out of time delayed elements of the known data. For simplicity, let us assume the known data to consist of a single element $u_i$ at any given discrete time $i$. Then the reconstructed phase space vector is defined as
\begin{equation}
    \vec{x}_i = \left(u_{i+\left(j-1\right)\tau}\right)_{j=1}^m ~,
\end{equation}
where $m$ and $\tau$ are parameters called the embedding dimension and time delay, respectively.

This method of phase space reconstruction is known to faithfully reproduce the full dynamics of the system. In fact, it has been proven that there exists a diffeomorphism between the original and reconstructed phase spaces~\cite{Takens:1981emb}.

\subsection{Model architecture}

\begin{figure}
    \centering
    \includegraphics[width=\linewidth]{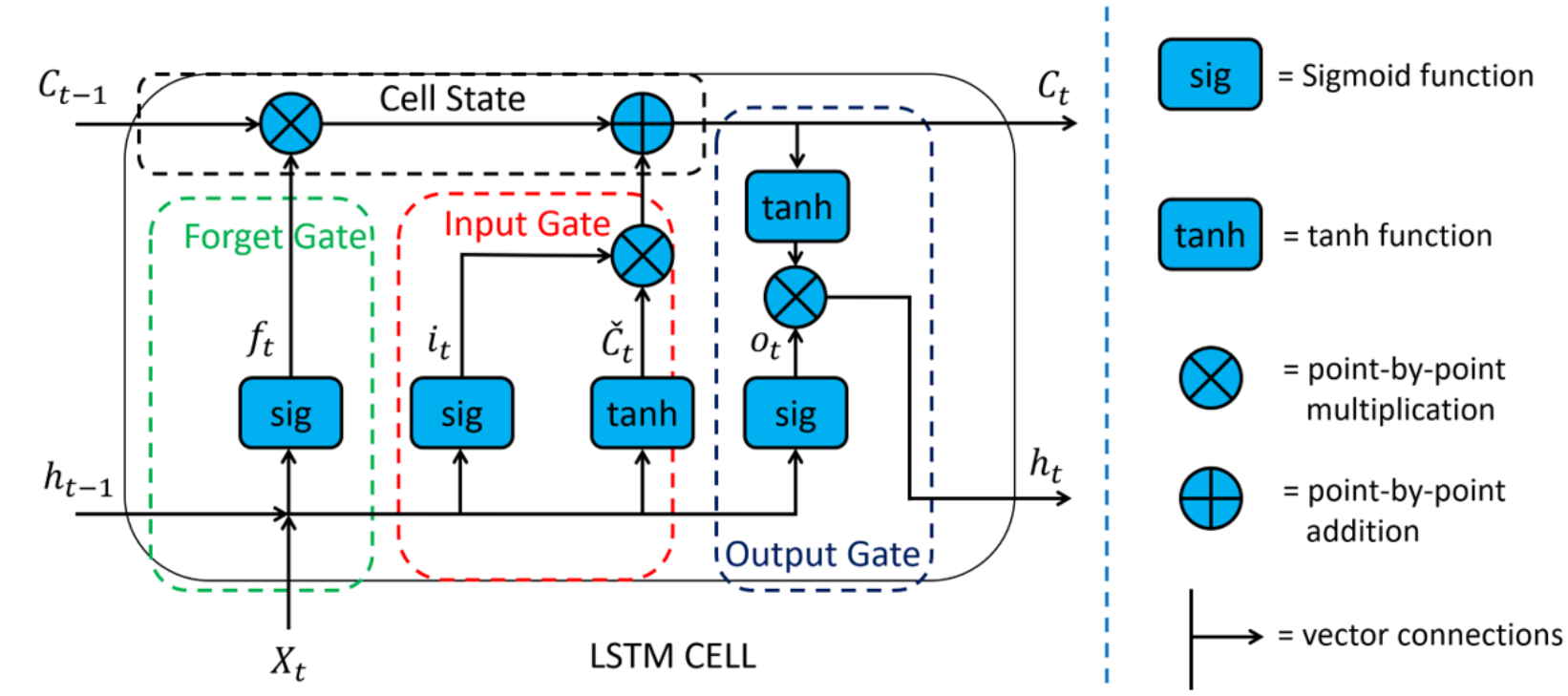}
    \caption{\Gls{lstm} network architecture. Taken from~\cite{Alfaidi:2024ioo}.}\label{fig:lstm}
\end{figure}

A \gls{lstm} network is built of an arbitrary number of \gls{lstm} cells, one of which is depicted in Fig.~\ref{fig:lstm}. They are specifically designed to handle time-series data: $X_t$ serves as the input and $o_t$ as the output at time $t$. The individual \gls{lstm} cells communicate through the short-term memory $h_t$ and the long-term memory $C_t$. For more details on \gls{lstm} architectures, see~\cite{Hochreiter:1997lstm}.

While in the original motivation the \gls{lstm} cells are stacked along the time direction of a time series, we follow a different approach: time-related properties of the system's orbits are encoded in the \gls{rqa} quantifiers, and we stack the \gls{lstm} cells in the direction along which we pick initial conditions. This allows our network to accept arbitrary-length inputs and return series with the same length. Furthermore, we apply a bidirectional \gls{lstm} architecture, which employs the same type of cells as shown in Fig.~\ref{fig:lstm}, in two sets: one running in each direction.

The network depth is kept as a parameter which we vary to determine an optimal value. Furthermore, dropout layers are included inside the \gls{lstm} network as well as before the output layer with a variable dropout rate which is also optimized later.

To build an input data sample, let us have a sequence of $N$ initial conditions, each producing a trajectory of a fixed length of 10000 points of a discrete-time system. The phase space is shifted and rescaled such that each dimension covers the interval $\left[0,\,2\pi\right]$. For each trajectory, we compute RR, DET, LAM, $L$, $L_\mathrm{entr}$, DIV, and $V_\mathrm{max}$ at the threshold values $\epsilon = 0.001$, 0.002, 0.005, 0.01, 0.02, 0.05, 0.1, 0.2, 0.5, 1. This produces a total of 70 \gls{rqa} indicators for each orbit. The network accepts as its input tensors of shape $N\times 70$ and produces an output in the shape $N\times 1$, meant to correspond to the network's degree of confidence that the corresponding orbit is part of a resonant island of stability, in the range $\left(0,\,1\right)$.

The network is composed of the \verb|torch.nn.LSTM| module with input and hidden sizes set to 70, followed by a dropout layer, a fully connected layer combining its input to a single output, and finally a Sigmoid layer maps it to the interval $\left[0,\,1\right]$.

Each dataset sample is a collection of these 70 recurrence quantifiers for a set of different trajectories of the system and represents a sweep of initial conditions along a line in the phase space.

\subsection{Training and validation data}

For training and validation data, we make use of a simple system typical for demonstrations of dynamical system properties. This is the Standard map \cite{Chirikov:1971gli} defined as
\begin{subequations}
\begin{align}
    x_{n+1} &= x_n + y_{n+1} ~,\\
    y_{n+1} &= y_n + K\sin\left(x_n\right) ~,
\end{align}\label{eqs:smap}
\end{subequations}
with $K$ a perturbation parameter. We have chosen the Standard map, since it is essentially a kicked pendulum and the dynamics of a pendulum essentially encapsulates the dynamics of a resonance \cite{Morbidelli02}.  Each dataset sample is generated for a fixed value of $K$ using a set of 1001 equidistant initial conditions covering the interval $y\in \left[-\pi,\, \pi\right]$ at $x = 0$. In particular, the training dataset consists of $K=0, 0.1,\dots 1$, while the validation dataset $K=0.05, 0.15,\dots 0.95$.

The data are hand-labeled: each initial condition is assigned the value 1 if it is part of a resonant island of stability, and 0 if it is not. As the purpose of this model is to detect the resonances based on correlations between recurrence quantifiers of different initial conditions, we only mark islands as resonant if there are at least two neighboring initial conditions in the island.

\subsection{Test data}

To demonstrate the effectiveness of the trained networks, we use test data coming from three different systems with specific properties.

\subsubsection{Dynamical mapping: de Vogeleare map}

For the simpler test data, we use the de Vogeleare map \cite{reichl2004}, defined as
\begin{subequations}
\begin{align}
    x_{n+1} &= - y_n + Kx_n + x_n^2 ~,\\
    y_{n+1} &= x_n - Kx_{n+1} - x_{n+1}^2 ~,
\end{align}
\end{subequations}
with $K\in\mathbb{R}$ a perturbation parameter.

\subsubsection{Poincar\'{e} section: motion around a black hole}\label{sssec:data_johpsa}

One of the aims of the method developed in this paper is the search for extended resonances in the motion of a test particle around a black hole. Knowledge of their locations will allow us to properly model the qualitatively distinct behavior of the passage through these resonances.

We consider a particle moving along a geodesic in the vicinity of a perturbed black hole, which could model the evolution of an \gls{emri} in modified gravity. The Johannsen-Psaltis spacetime is given by the line element~\cite{Johannsen2011PhRvD, Zelenka:2017aqn}
\begin{equation}
\mathrm{d}s^2 = g_{tt}\mathrm{d}t^2 + g_{rr}\mathrm{d}r^2 + g_{\theta\theta}\mathrm{d}\theta^2 + g_{\phi\phi}\mathrm{d}\phi^2 + 2g_{t\phi}\mathrm{d}t\mathrm{d}\phi
\end{equation}
with metric components
\begin{subequations}
\begin{align}
    g_{tt} &= -\left(1+h\right)\left(1 - \frac{2Mr}{\Sigma}\right) ~,\\
    g_{t\phi} &= -\frac{2aMr\sin^2\theta}{\Sigma}\left(1+h\right) ~,\\
    g_{\phi\phi} &= \frac{\Lambda\sin^2\theta}{\Sigma} + ha^2\left(1 + \frac{2Mr}{\Sigma}\right)\sin^4\theta ~,\\
    g_{rr} &= \frac{\Sigma\left(1+h\right)}{\Delta + a^2h\sin^2\theta} ~,\\
    g_{\theta\theta} &= \Sigma ~.
\end{align}
\end{subequations}
using the functions
\begin{subequations}
\begin{align}
    \Sigma &= r^2 + a^2\cos^2\theta ~,\\
    h &= \sum_{k=0}^\infty \left(\epsilon_{2k}+\epsilon_{2k+1}\frac{Mr}{\Sigma}\right)\left(\frac{M^2}{\Sigma}\right)^k ~,\\
    \Delta &= r^2 + a^2 - 2Mr ~,\\
    \omega^2 &= r^2 + a^2 ~,\\
    \Lambda &= \omega^4 - a^2\Delta\sin^2\theta ~.
\end{align}
\end{subequations}

We do not go into detail on properties of geodesic motion in the Johannsen-Psaltis spacetime; for a summary, see~\cite{Zelenka:2017aqn}. Points on the Poincar\'{e} section given by the equatorial plane $\theta=\pi/2$, $\dot{\theta}>0$ parametrized by $r,\, p^r$ are used for recurrence analysis. In this system, points from the outermost orbits which reach near the tip of the main island of stability are highly concentrated in this region, with much lower densities at higher values of $r$. This affects the \gls{rqa} quantifiers and is therefore a good test case for the present \gls{lstm} networks.

Following~\cite{Zelenka:2017aqn}, we choose the spacetime parameter values $a=0.5M$, $\epsilon_3=0.3$, and $\epsilon_k = 0$ for any $k\neq 3$. For integrals of motion we choose the values $E = 0.95\mu$, $L_z = 2.85\mu M$. The initial conditions are taken along the $p^r=0$ line on the Poincar\'{e} section in $r\in\left[6.36M,\, 6.43M\right]$ spaced at $0.001M$. This sweep of initial conditions includes the $\omega_{r}/\omega_{\theta} = 2/3$ orbital resonance.

\subsubsection{Higher-dimensional example: 4D map}\label{sssec:data_4d_map}

In a Hamiltonian system with 2 degrees of freedom, alternative methods exist which can detect resonances, such as the rotation curve and its estimation using the 2-dimensional Poincar\'{e} section. However, already in a system with 3 degrees of freedom, the Poincar\'{e} section becomes 4-dimensional and these simple methods are no longer applicable.

As a result, one of the goals of the method developed in this article is to allow detection of resonances in higher-dimensional systems. Since the recurrence plots and quantifiers are fairly universal measures of the system's properties, we apply the model trained on 2-dimensional data to 4-dimensional data generated by a map defined as~\cite{LukesGerakopoulos2008:4dmap, Froeschl2005}
\begin{subequations}
\begin{align}
    x_{n+1} &= x_n - K\frac{\sin\left(x_n + y_n\right)}{\left(\cos\left(x_n + y_n\right) + \cos\left(z_n + t_n\right) + 4\right)^2} ~,\\
    y_{n+1} &= y_n + x_n \quad\quad\mathrm{mod}~2\pi ~,\\
    z_{n+1} &= x_n - K\frac{\sin\left(z_n + t_n\right)}{\left(\cos\left(x_n + y_n\right) + \cos\left(z_n + t_n\right) + 4\right)^2} ~,\\
    t_{n+1} &= t_n + z_n \quad\quad\mathrm{mod}~2\pi ~.
\end{align}\label{eqs:4d_map}
\end{subequations}
The $x$ and $z$ coordinates correspond to actions, and $y$ and $t$ angles in a 2-degree-of-freedom map, and $K$ is a perturbation parameter (the standard notation would be $\epsilon$~\cite{LukesGerakopoulos2008:4dmap}, which would clash with the recurrence threshold in this work).

\begin{figure}
    \centering
    \includegraphics[width=\linewidth]{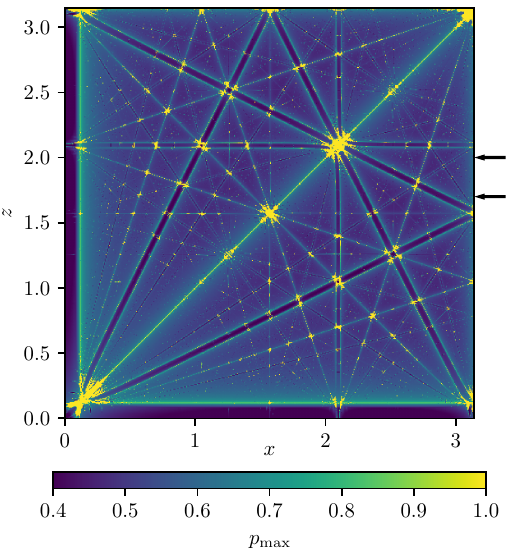}
    \caption{Geography of the resonances of the 4D map in the $x$, $z$ plane. The arrows on the right side show the $z$ values corresponding to the initial conditions taken in Figs.~\ref{fig:test_4d_map} and~\ref{fig:test_4d_map_embed}. The choice of the color mapping between 0.4 and 1 precludes us from identifying regular and chaotic trajectories individually; however, it clearly shows thin chaotic layers at the edges of resonances. The APLE is computed on a $501\times 501$ grid and the perturbation parameter is set to $K=0.05$.}
    \label{fig:aple_4d_map}
\end{figure}

While the full structure of this map's phase space is difficult to understand, we can visualize the resonances in the space of the actions (corresponding to angular frequencies) $x$ and $z$ using the chaotic indicator APLE, defined as
\begin{equation}
    \mathrm{APLE} = \sup_{t_1 < t \leq T}\left(\frac{\log\left(\xi\left(t\right)/\xi\left(t_1\right)\right)}{\log\left(t/t_1\right)}\right) ~.
\end{equation}
This expression converges to 1 or diverges when $\xi\left(t\right)$ is a linear or exponential function, respectively. Therefore 1 is frequently chosen as a threshold to distinguish regular from chaotic trajectories. The geography of resonances in the 4D map of Eq.~\eqref{eqs:4d_map} is shown in Fig.~\ref{fig:aple_4d_map}.

We use two sweeps of initial conditions along $x\in\left[0, 2\pi\right]$ at $z=1.7$ and $z=2$, respectively. The grid along the $x$ axis consists of 4001 samples with uniform spacing over the whole interval. All initial conditions satisfy $y=t=0$. Following Fig.~\ref{fig:aple_4d_map}, the perturbation parameter is set to $K=0.05$ in both cases.

\section{Results} \label{sec:Results}

We train the networks for 1000 epochs using the Adam optimizer~\cite{Adam} using a learning rate of $\eta = 10^{-5}$ and the PyTorch default values $\beta_1 = 0.9$, $\beta_2 = 0.999$, $\varepsilon = 10^{-8}$. The loss used for optimization is the mean squared error loss
\begin{subequations}
    \begin{align}
        &\mathrm{MSE}:~\mathbb{R}^{mn}\times\mathbb{R}^{mn} \to \mathbb{R} ~,\\
        &\bar{\mathbf{Y}},\,\mathbf{Y} \mapsto \frac{1}{mn}\sum_{i=1}^m \sum_{j=1}^n \left(\bar{Y}_{ij} - Y_{ij}\right)^2 ~.
    \end{align}
\end{subequations}
Early experiments show that this produces much better results than using a binary cross entropy which seems the obvious choice for what is in principle a binary classification problem.

The network is trained on a grid of integer depths from 1 to 10 and dropout rates of 0, 0.1, 0.2, 0.3, and 0.5. At each grid point, 4 independent networks are initialized and trained to prevent corruption due to insufficient sampling of the parameter space.

\subsection{Basic network}

\begin{figure}
    \centering
    \includegraphics[width=\linewidth]{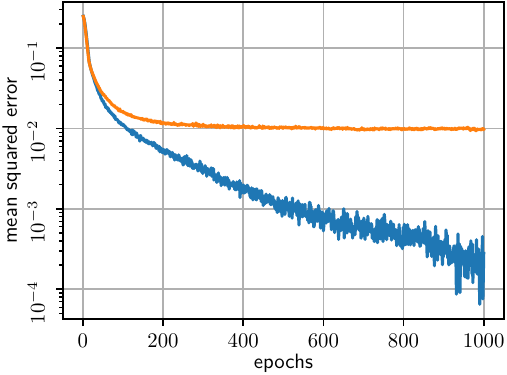}
    \caption{Evolution of the training (blue) and validation (orange) losses. Initial conditions are taken on the $y=0$ line. }
    \label{fig:loss_curves}
\end{figure}

The network which reached the lowest validation loss is one of the networks with 2 layers and dropout rate set to 0.5, its loss evolution is shown in Fig.~\ref{fig:loss_curves}.

\subsubsection{de Vogeleare map}\label{ssec:test_devog}

\begin{figure*}
    \centering
    \includegraphics[width=\linewidth]{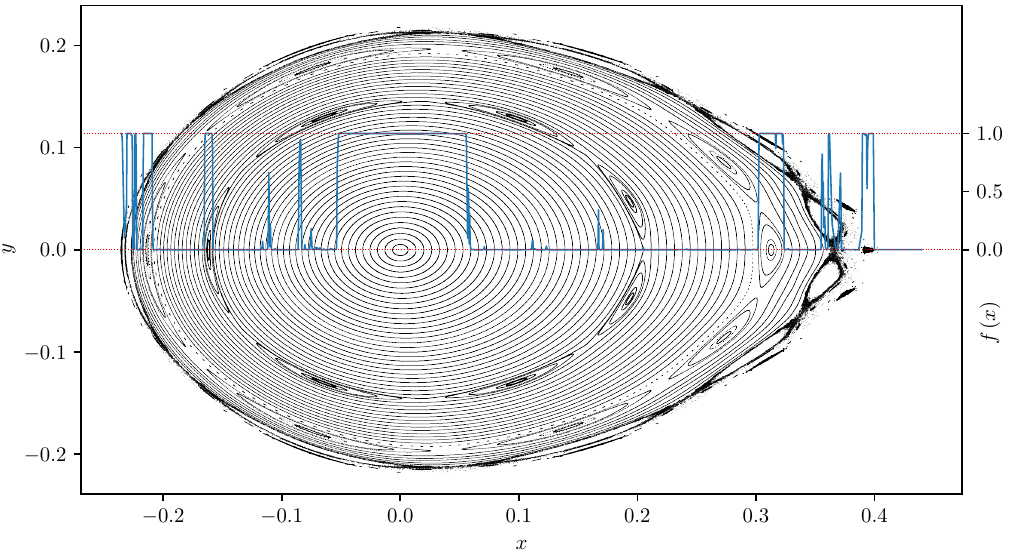}
    \caption{Results on the de Vogeleare map. The Poincar\'{e} section is shown in black, displaying several prominent resonances. The blue curve displays the outputs of the neural network from inputs based on initial conditions along the $x$ axis with $y=0$, the dotted red lines corresponding to the values 0 and 1 of the neural network output. While the network output is based on a total of 2001 initial conditions, the Poincar\'{e} section only uses 96 initial conditions to improve the clarity of the plot.}
    \label{fig:test_devog}
\end{figure*}

To demonstrate the effectiveness and robustness of the resonance detection network, we apply it to test data generated by a different system than the standard map. Fig.~\ref{fig:test_devog} shows the results of testing on the de Vogeleare map data. The perturbation parameter is set to $K=0.56$ and 2001 initial conditions are used with linear spacing in $x\in\left[-0.44,\,0.44\right]$ and $y=0$. Initial conditions which lead to escaping trajectories are excluded on the basis of either coordinate's absolute value exceeding $10^4$ at any point of the evolution.

All resonances in Fig.~\ref{fig:test_devog} are detected clearly by the network, with only few spurious jumps inside the main island of stability and in the chaotic zone on its right side. Besides these, the center of the main island of stability is also marked as a resonance. 

\subsubsection{Motion around a black hole}

\begin{figure*}
    \centering
    \includegraphics[width=\linewidth]{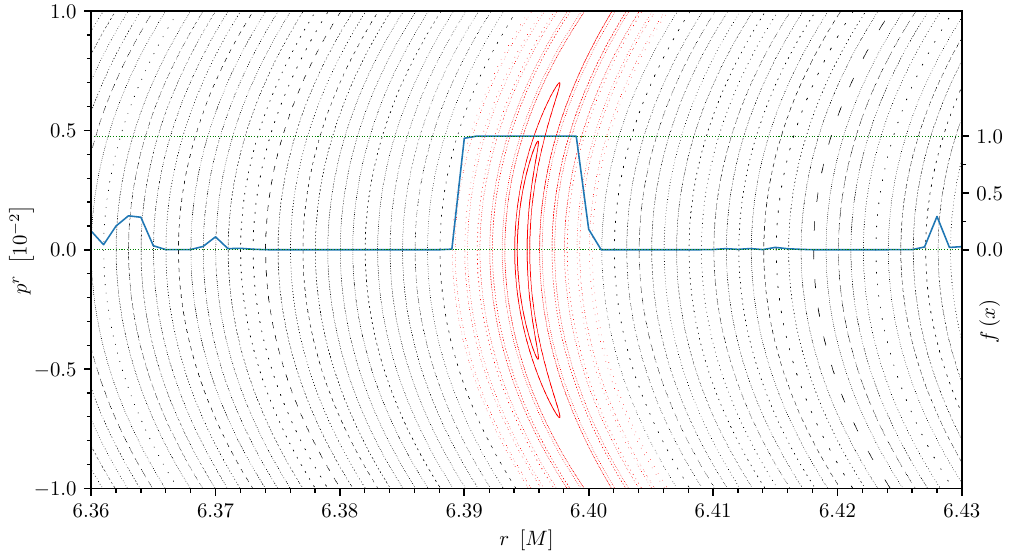}
    \caption{Results of the basic network on the Poincar\'{e} section of geodesic motion in the Johannsen-Psaltis spacetime superimposed on the corresponding Poincar\'{e} section. Shown in black and red are successive intersections of orbits corresponding to initial conditions along the $p^r=0$ line with initial $r$ between 6.35 and 6.43, spaced at 0.001. 10000 intersections are shown per initial condition, and the 1/3 resonance is highlighted in red, while the KAM orbits are in black. The blue line shows the outputs of the neural network as a function of the intitial condition along the $p^r=0$ line, with green dotted lines showing the edge values of 0 and 1.}\label{fig:test_johpsa}
\end{figure*}

Fig.~\ref{fig:test_johpsa} shows the result of the basic network applied to the Poincar\'{e} section of a test particle following a geodesic in the Johannsen-Psaltis spacetime metric (see Sec.~\ref{sssec:data_johpsa}). The section contains a single resonance $\omega_r/\omega_\theta = 2/3$, which is detected by the network very clearly. We note that while the figure shows surface of section points with initial starting at initial $r=6.35M$ in order to fill out the top and bottom left corners of the plot, the network analyzes data starting at initial $r=6.36M$.

\subsubsection{4D map}

\begin{figure*}
    \centering
    \includegraphics[width=\linewidth]{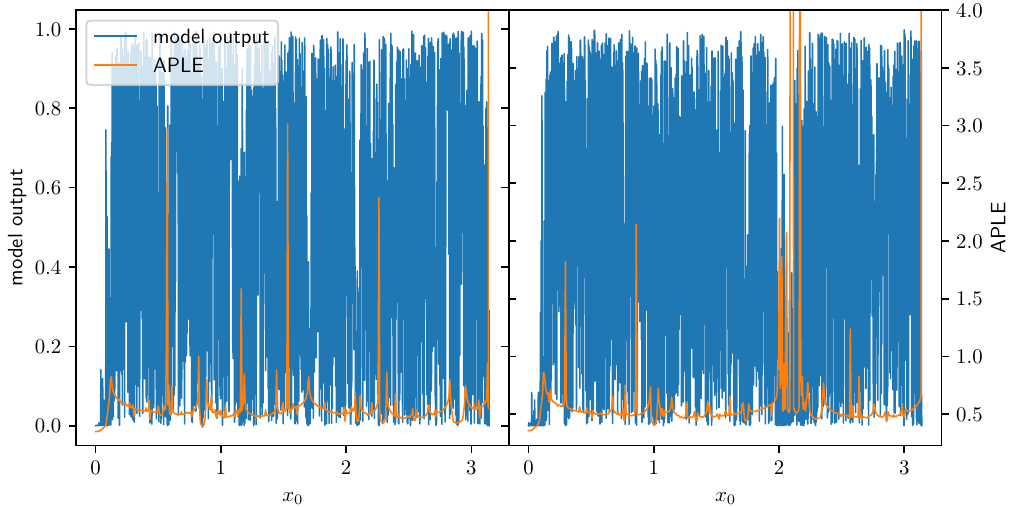}
    \caption{Results of the basic network on the 4D map. Each panel corresponds to 4001 initial conditions with $x_0\in\left[0,\,\pi\right]$ and linear spacing, and $y_0 = t_0 = 0$. In the left panel, it holds $z_0 = 1.7$, and in the right, $z_0 = 2$. The primary $y$ axis on the left corresponds to both model output curves, and the secondary $y$ axis on the right corresponds to both APLE curves.}
    \label{fig:test_4d_map}
\end{figure*}

Finally, let us focus on the 4-dimensional map. Fig.~\ref{fig:test_4d_map} shows the output of the basic network over the two curves described in Sec.~\ref{sssec:data_4d_map}. For better orientation, the corresponding APLE curves are also shown.

In this case, no resonances are clearly recognizable in the network output.

\subsection{Embedded network}

While the full phase space contains more information than its reconstruction through embedding, there are cases where recurrence analysis may be more effective when applied to embedded data.

We train the same network architecture as the basic network with only one modification: the training recurrence quantifiers are recomputed based on the same original trajectories but only using the $x$ coordinate with embedding. We use embedding dimension $m=2$ and time delay $\tau = 1$ for the embedded training data. These parameters have been determined using the false nearest neighbor algorithm and as minimum of the mutual information, respectively, on several trajectories representative of the dataset as a whole~\cite{Sukova:2015naa}. Again, the lowest validation loss is reached by one of the networks with 2 layers and a dropout rate of 0.5.

\begin{figure*}
    \centering
    \includegraphics[width=\linewidth]{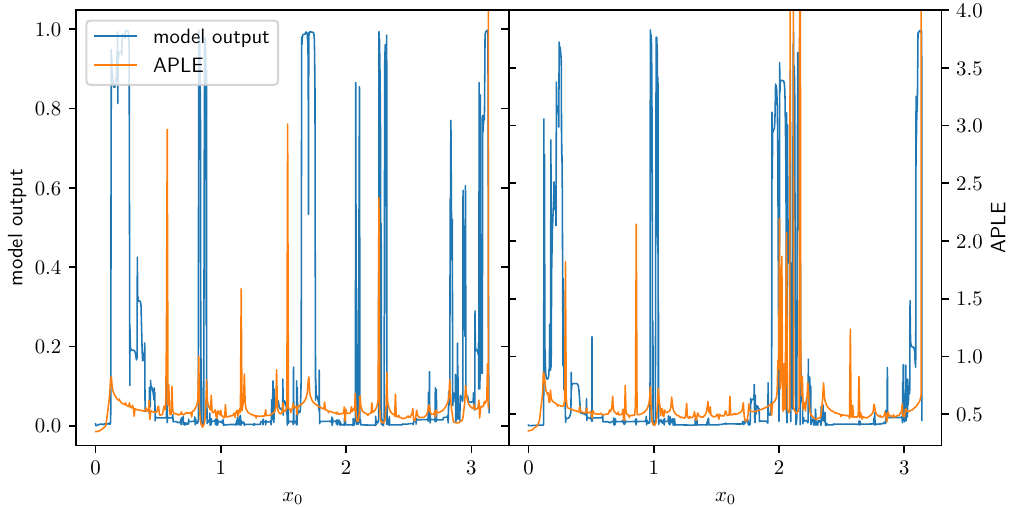}
    \caption{Results of the embedded network on the 4D map. Each panel corresponds to 4001 initial conditions with $x_0\in\left[0,\,\pi\right]$ and linear spacing, and $y_0 = t_0 = 0$. In the left panel, it holds $z_0 = 1.7$, and in the right, $z_0 = 2$. The primary $y$ axis on the left corresponds to both model output curves, and the secondary $y$ axis on the right corresponds to both APLE curves.}
    \label{fig:test_4d_map_embed}
\end{figure*}

We perform the same experiment as in Fig.~\ref{fig:test_4d_map} using the embedded network. To compute recurrence quantifiers of the 4D map data, we use embedding with dimension $m=4$ and time delay $\tau = 2$, set using the same algorithms as for the training data. Fig.~\ref{fig:test_4d_map_embed} shows the results of the experiment. The classification is not as clear as, e.g., in the case of the 2D mapping (see Sec.~\ref{ssec:test_devog}). At the same time, the network is reacting to resonances in their true locations, albeit with variable heights of the output peaks as opposed to the 2D case.

We have also attempted to apply the embedded network to non-embedded 4D map data, as well as the basic network to embedded 4D map data. In both cases, the results were similar to those of Fig.~\ref{fig:test_4d_map} and no resonances were recognizable. Hence, the only combination that works is to use embedded network to embedded 4D map data.

\section{Conclusion}\label{sec:Conc}

In this work, we have developed a machine-learning based method to uncover orbital resonances using \acrfull{rqa} in dynamical maps. Using the Poincar\'{e} section method, applications to continuous-time data follow in a straightforward manner.

We have demonstrated the effectiveness of our algorithm on two primary test cases:
\begin{itemize}
    \item the Poincar\'{e} section of a test particle near a perturbed black hole, as a model for the conservative portion of an \gls{emri}, whose evolution is highly non-homogeneous in the phase space,
    \item a dynamical map in a 4-dimensional phase space, where the visualization tools which are critical for conventional methods are absent.
\end{itemize}

While the identification of resonances in the 4-dimensional case is not as clear as in 2-dimensional cases, we conclude that this method is effective. Yet, there are multiple network parameters which can be optimized, along with engineering more general training data which may improve the performance of the method.

One of the primary motivations for this research is to identify orbital resonances due to the spin of the secondary in an \gls{emri}. In this case, both issues above are relevant, as the general shape of the phase space remains similar in the orbital degrees of freedom, and another degree of freedom is added when the secondary gains a spin.

At the same time, Birkhoff chains arise generically in near-integrable systems, and appear in a large portion of problems in physics and related fields. As such, methods which aid in their detection and proper treatment are critical in many problems in physics and related fields.

\section{Acknowledgments}

The authors are supported by the fellowship Lumina Quaeruntur No. LQ100032102 of the Czech Academy of Sciences.

\bibliographystyle{apsrev4-2}
\bibliography{bibliography}

\begin{thebibliography}{31}%
\makeatletter
\providecommand \@ifxundefined [1]{%
 \@ifx{#1\undefined}
}%
\providecommand \@ifnum [1]{%
 \ifnum #1\expandafter \@firstoftwo
 \else \expandafter \@secondoftwo
 \fi
}%
\providecommand \@ifx [1]{%
 \ifx #1\expandafter \@firstoftwo
 \else \expandafter \@secondoftwo
 \fi
}%
\providecommand \natexlab [1]{#1}%
\providecommand \enquote  [1]{``#1''}%
\providecommand \bibnamefont  [1]{#1}%
\providecommand \bibfnamefont [1]{#1}%
\providecommand \citenamefont [1]{#1}%
\providecommand \href@noop [0]{\@secondoftwo}%
\providecommand \href [0]{\begingroup \@sanitize@url \@href}%
\providecommand \@href[1]{\@@startlink{#1}\@@href}%
\providecommand \@@href[1]{\endgroup#1\@@endlink}%
\providecommand \@sanitize@url [0]{\catcode `\\12\catcode `\$12\catcode
  `\&12\catcode `\#12\catcode `\^12\catcode `\_12\catcode `\%12\relax}%
\providecommand \@@startlink[1]{}%
\providecommand \@@endlink[0]{}%
\providecommand \url  [0]{\begingroup\@sanitize@url \@url }%
\providecommand \@url [1]{\endgroup\@href {#1}{\urlprefix }}%
\providecommand \urlprefix  [0]{URL }%
\providecommand \Eprint [0]{\href }%
\providecommand \doibase [0]{https://doi.org/}%
\providecommand \selectlanguage [0]{\@gobble}%
\providecommand \bibinfo  [0]{\@secondoftwo}%
\providecommand \bibfield  [0]{\@secondoftwo}%
\providecommand \translation [1]{[#1]}%
\providecommand \BibitemOpen [0]{}%
\providecommand \bibitemStop [0]{}%
\providecommand \bibitemNoStop [0]{.\EOS\space}%
\providecommand \EOS [0]{\spacefactor3000\relax}%
\providecommand \BibitemShut  [1]{\csname bibitem#1\endcsname}%
\let\auto@bib@innerbib\@empty
\bibitem [{\citenamefont {Strogatz}(2019)}]{strogatz19}%
  \BibitemOpen
  \bibfield  {author} {\bibinfo {author} {\bibfnamefont {S.~H.}\ \bibnamefont
  {Strogatz}},\ }\href@noop {} {\emph {\bibinfo {title} {Nonlinear Dynamics and
  Chaos}}},\ \bibinfo {edition} {2nd}\ ed.\ (\bibinfo  {publisher} {CRC
  Press},\ \bibinfo {address} {London, England},\ \bibinfo {year}
  {2019})\BibitemShut {NoStop}%
\bibitem [{\citenamefont {{Mukherjee}}\ \emph {et~al.}(2023)\citenamefont
  {{Mukherjee}}, \citenamefont {{Kop{\'a}{\v{c}}ek}},\ and\ \citenamefont
  {{Lukes-Gerakopoulos}}}]{mukherjee23}%
  \BibitemOpen
  \bibfield  {author} {\bibinfo {author} {\bibfnamefont {S.}~\bibnamefont
  {{Mukherjee}}}, \bibinfo {author} {\bibfnamefont {O.}~\bibnamefont
  {{Kop{\'a}{\v{c}}ek}}},\ and\ \bibinfo {author} {\bibfnamefont
  {G.}~\bibnamefont {{Lukes-Gerakopoulos}}},\ }\href
  {https://doi.org/10.1103/PhysRevD.107.064005} {\bibfield  {journal} {\bibinfo
   {journal} {\prd}\ }\textbf {\bibinfo {volume} {107}},\ \bibinfo {eid}
  {064005} (\bibinfo {year} {2023})},\ \Eprint
  {https://arxiv.org/abs/2206.10302} {arXiv:2206.10302 [gr-qc]} \BibitemShut
  {NoStop}%
\bibitem [{\citenamefont {Zelenka}\ \emph {et~al.}(2020)\citenamefont
  {Zelenka}, \citenamefont {Lukes-Gerakopoulos}, \citenamefont {Witzany},\ and\
  \citenamefont {Kop\'a\v{c}ek}}]{Zelenka:2019nyp}%
  \BibitemOpen
  \bibfield  {author} {\bibinfo {author} {\bibfnamefont {O.}~\bibnamefont
  {Zelenka}}, \bibinfo {author} {\bibfnamefont {G.}~\bibnamefont
  {Lukes-Gerakopoulos}}, \bibinfo {author} {\bibfnamefont {V.}~\bibnamefont
  {Witzany}},\ and\ \bibinfo {author} {\bibfnamefont {O.}~\bibnamefont
  {Kop\'a\v{c}ek}},\ }\href {https://doi.org/10.1103/PhysRevD.101.024037}
  {\bibfield  {journal} {\bibinfo  {journal} {Phys. Rev. D}\ }\textbf {\bibinfo
  {volume} {101}},\ \bibinfo {pages} {024037} (\bibinfo {year}
  {2020})}\BibitemShut {NoStop}%
\bibitem [{\citenamefont {{Amaro-Seoane}}(2018)}]{amaro18}%
  \BibitemOpen
  \bibfield  {author} {\bibinfo {author} {\bibfnamefont {P.}~\bibnamefont
  {{Amaro-Seoane}}},\ }\href {https://doi.org/10.1007/s41114-018-0013-8}
  {\bibfield  {journal} {\bibinfo  {journal} {Living Reviews in Relativity}\
  }\textbf {\bibinfo {volume} {21}},\ \bibinfo {eid} {4} (\bibinfo {year}
  {2018})},\ \Eprint {https://arxiv.org/abs/1205.5240} {arXiv:1205.5240
  [astro-ph.CO]} \BibitemShut {NoStop}%
\bibitem [{\citenamefont {{Amaro-Seoane}}\ \emph {et~al.}(2023)\citenamefont
  {{Amaro-Seoane}}, \citenamefont {{Andrews}}, \citenamefont {{Arca Sedda}},
  \citenamefont {{Askar}}, \citenamefont {{Baghi}}, \citenamefont {{Balasov}},
  \citenamefont {{Bartos}}, \citenamefont {{Bavera}}, \citenamefont
  {{Bellovary}}, \citenamefont {{Berry}}, \citenamefont {{Berti}},
  \citenamefont {{Bianchi}}, \citenamefont {{Blecha}}, \citenamefont
  {{Blondin}}, \citenamefont {{Bogdanovi{\'c}}} \emph {et~al.}}]{amaro23}%
  \BibitemOpen
  \bibfield  {author} {\bibinfo {author} {\bibfnamefont {P.}~\bibnamefont
  {{Amaro-Seoane}}}, \bibinfo {author} {\bibfnamefont {J.}~\bibnamefont
  {{Andrews}}}, \bibinfo {author} {\bibfnamefont {M.}~\bibnamefont {{Arca
  Sedda}}}, \bibinfo {author} {\bibfnamefont {A.}~\bibnamefont {{Askar}}},
  \bibinfo {author} {\bibfnamefont {Q.}~\bibnamefont {{Baghi}}}, \bibinfo
  {author} {\bibfnamefont {R.}~\bibnamefont {{Balasov}}}, \bibinfo {author}
  {\bibfnamefont {I.}~\bibnamefont {{Bartos}}}, \bibinfo {author}
  {\bibfnamefont {S.~S.}\ \bibnamefont {{Bavera}}}, \bibinfo {author}
  {\bibfnamefont {J.}~\bibnamefont {{Bellovary}}}, \bibinfo {author}
  {\bibfnamefont {C.~P.~L.}\ \bibnamefont {{Berry}}}, \bibinfo {author}
  {\bibfnamefont {E.}~\bibnamefont {{Berti}}}, \bibinfo {author} {\bibfnamefont
  {S.}~\bibnamefont {{Bianchi}}}, \bibinfo {author} {\bibfnamefont
  {L.}~\bibnamefont {{Blecha}}}, \bibinfo {author} {\bibfnamefont
  {S.}~\bibnamefont {{Blondin}}}, \bibinfo {author} {\bibfnamefont
  {T.}~\bibnamefont {{Bogdanovi{\'c}}}}, \emph {et~al.},\ }\href
  {https://doi.org/10.1007/s41114-022-00041-y} {\bibfield  {journal} {\bibinfo
  {journal} {Living Reviews in Relativity}\ }\textbf {\bibinfo {volume} {26}},\
  \bibinfo {eid} {2} (\bibinfo {year} {2023})},\ \Eprint
  {https://arxiv.org/abs/2203.06016} {arXiv:2203.06016 [gr-qc]} \BibitemShut
  {NoStop}%
\bibitem [{\citenamefont {{Lukes-Gerakopoulos}}\ and\ \citenamefont
  {{Witzany}}(2021)}]{LG:2021hgwa}%
  \BibitemOpen
  \bibfield  {author} {\bibinfo {author} {\bibfnamefont {G.}~\bibnamefont
  {{Lukes-Gerakopoulos}}}\ and\ \bibinfo {author} {\bibfnamefont
  {V.}~\bibnamefont {{Witzany}}},\ }in\ \href
  {https://doi.org/10.1007/978-981-15-4702-7_42-1} {\emph {\bibinfo {booktitle}
  {Handbook of Gravitational Wave Astronomy}}},\ \bibinfo {editor} {edited by\
  \bibinfo {editor} {\bibfnamefont {C.}~\bibnamefont {{Bambi}}}, \bibinfo
  {editor} {\bibfnamefont {S.}~\bibnamefont {{Katsanevas}}},\ and\ \bibinfo
  {editor} {\bibfnamefont {K.~D.}\ \bibnamefont {{Kokkotas}}}}\ (\bibinfo
  {year} {2021})\ p.~\bibinfo {pages} {42}\BibitemShut {NoStop}%
\bibitem [{\citenamefont {{Sukov{\'a}}}\ and\ \citenamefont
  {{Semer{\'a}k}}(2013)}]{Sukova:2013MNRAS}%
  \BibitemOpen
  \bibfield  {author} {\bibinfo {author} {\bibfnamefont {P.}~\bibnamefont
  {{Sukov{\'a}}}}\ and\ \bibinfo {author} {\bibfnamefont {O.}~\bibnamefont
  {{Semer{\'a}k}}},\ }\href {https://doi.org/10.1093/mnras/stt1587} {\bibfield
  {journal} {\bibinfo  {journal} {\mnras}\ }\textbf {\bibinfo {volume} {436}},\
  \bibinfo {pages} {978} (\bibinfo {year} {2013})},\ \Eprint
  {https://arxiv.org/abs/1308.4306} {arXiv:1308.4306 [gr-qc]} \BibitemShut
  {NoStop}%
\bibitem [{\citenamefont {{Hartl}}(2003)}]{hartl03}%
  \BibitemOpen
  \bibfield  {author} {\bibinfo {author} {\bibfnamefont {M.~D.}\ \bibnamefont
  {{Hartl}}},\ }\href {https://doi.org/10.1103/PhysRevD.67.024005} {\bibfield
  {journal} {\bibinfo  {journal} {\prd}\ }\textbf {\bibinfo {volume} {67}},\
  \bibinfo {eid} {024005} (\bibinfo {year} {2003})},\ \Eprint
  {https://arxiv.org/abs/gr-qc/0210042} {arXiv:gr-qc/0210042 [gr-qc]}
  \BibitemShut {NoStop}%
\bibitem [{\citenamefont {Marwan}\ \emph {et~al.}(2007)\citenamefont {Marwan},
  \citenamefont {{Carmen Romano}}, \citenamefont {Thiel},\ and\ \citenamefont
  {Kurths}}]{Marwan:2007rps}%
  \BibitemOpen
  \bibfield  {author} {\bibinfo {author} {\bibfnamefont {N.}~\bibnamefont
  {Marwan}}, \bibinfo {author} {\bibfnamefont {M.}~\bibnamefont {{Carmen
  Romano}}}, \bibinfo {author} {\bibfnamefont {M.}~\bibnamefont {Thiel}},\ and\
  \bibinfo {author} {\bibfnamefont {J.}~\bibnamefont {Kurths}},\ }\href
  {https://doi.org/https://doi.org/10.1016/j.physrep.2006.11.001} {\bibfield
  {journal} {\bibinfo  {journal} {Physics Reports}\ }\textbf {\bibinfo {volume}
  {438}},\ \bibinfo {pages} {237} (\bibinfo {year} {2007})}\BibitemShut
  {NoStop}%
\bibitem [{\citenamefont {{Johannsen}}\ and\ \citenamefont
  {{Psaltis}}(2011)}]{Johannsen2011PhRvD}%
  \BibitemOpen
  \bibfield  {author} {\bibinfo {author} {\bibfnamefont {T.}~\bibnamefont
  {{Johannsen}}}\ and\ \bibinfo {author} {\bibfnamefont {D.}~\bibnamefont
  {{Psaltis}}},\ }\href {https://doi.org/10.1103/PhysRevD.83.124015} {\bibfield
   {journal} {\bibinfo  {journal} {\prd}\ }\textbf {\bibinfo {volume} {83}},\
  \bibinfo {eid} {124015} (\bibinfo {year} {2011})},\ \Eprint
  {https://arxiv.org/abs/1105.3191} {arXiv:1105.3191 [gr-qc]} \BibitemShut
  {NoStop}%
\bibitem [{\citenamefont {{Kolmogorov}}(1954)}]{Kolmogorov54}%
  \BibitemOpen
  \bibfield  {author} {\bibinfo {author} {\bibfnamefont {A.~N.}\ \bibnamefont
  {{Kolmogorov}}},\ }\href@noop {} {\bibfield  {journal} {\bibinfo  {journal}
  {Doklady Akademii Nauk SSSR}\ }\textbf {\bibinfo {volume} {98}},\ \bibinfo
  {pages} {527} (\bibinfo {year} {1954})}\BibitemShut {NoStop}%
\bibitem [{\citenamefont {{Arnold}}(1963)}]{Arnold63}%
  \BibitemOpen
  \bibfield  {author} {\bibinfo {author} {\bibfnamefont {V.~I.}\ \bibnamefont
  {{Arnold}}},\ }\href {https://doi.org/10.1070/RM1963v018n05ABEH004130}
  {\bibfield  {journal} {\bibinfo  {journal} {Russian Mathematical Surveys}\
  }\textbf {\bibinfo {volume} {18}},\ \bibinfo {pages} {9} (\bibinfo {year}
  {1963})}\BibitemShut {NoStop}%
\bibitem [{\citenamefont {{Moser}}(1962)}]{Moser62}%
  \BibitemOpen
  \bibfield  {author} {\bibinfo {author} {\bibfnamefont {J.}~\bibnamefont
  {{Moser}}},\ }\href@noop {} {\bibfield  {journal} {\bibinfo  {journal}
  {Nachrichten der Akademie der Wissenschaften in G\"ottingen. {II}.
  Mathematisch-Physikalische Klasse}\ ,\ \bibinfo {pages} {1}} (\bibinfo {year}
  {1962})}\BibitemShut {NoStop}%
\bibitem [{\citenamefont {{Birkhoff}}(1913)}]{Birkhoff13}%
  \BibitemOpen
  \bibfield  {author} {\bibinfo {author} {\bibfnamefont {G.~D.}\ \bibnamefont
  {{Birkhoff}}},\ }\href {https://doi.org/10.1090/S0002-9947-1913-1500933-9}
  {\bibfield  {journal} {\bibinfo  {journal} {Transactions of the American
  Mathematical Society}\ }\textbf {\bibinfo {volume} {14}},\ \bibinfo {pages}
  {14} (\bibinfo {year} {1913})}\BibitemShut {NoStop}%
\bibitem [{\citenamefont {{Marwan}}(2011)}]{marwan11}%
  \BibitemOpen
  \bibfield  {author} {\bibinfo {author} {\bibfnamefont {N.}~\bibnamefont
  {{Marwan}}},\ }\href {https://doi.org/10.1142/S0218127411029008} {\bibfield
  {journal} {\bibinfo  {journal} {International Journal of Bifurcation and
  Chaos}\ }\textbf {\bibinfo {volume} {21}},\ \bibinfo {pages} {1003} (\bibinfo
  {year} {2011})},\ \Eprint {https://arxiv.org/abs/1007.2215} {arXiv:1007.2215
  [nlin.CD]} \BibitemShut {NoStop}%
\bibitem [{\citenamefont {{Lukes-Gerakopoulos}}\ and\ \citenamefont
  {{Kop{\'a}{\v{c}}ek}}(2018)}]{glg18}%
  \BibitemOpen
  \bibfield  {author} {\bibinfo {author} {\bibfnamefont {G.}~\bibnamefont
  {{Lukes-Gerakopoulos}}}\ and\ \bibinfo {author} {\bibfnamefont
  {O.}~\bibnamefont {{Kop{\'a}{\v{c}}ek}}},\ }\href
  {https://doi.org/10.1142/S0218271818500104} {\bibfield  {journal} {\bibinfo
  {journal} {International Journal of Modern Physics D}\ }\textbf {\bibinfo
  {volume} {27}},\ \bibinfo {eid} {1850010} (\bibinfo {year} {2018})},\ \Eprint
  {https://arxiv.org/abs/1709.08446} {arXiv:1709.08446 [gr-qc]} \BibitemShut
  {NoStop}%
\bibitem [{\citenamefont {{Kop{\'a}{\v{c}}ek}}\ \emph
  {et~al.}(2010)\citenamefont {{Kop{\'a}{\v{c}}ek}}, \citenamefont {{Karas}},
  \citenamefont {{Kov{\'a}{\v{r}}}},\ and\ \citenamefont
  {{Stuchl{\'\i}k}}}]{kopacek10}%
  \BibitemOpen
  \bibfield  {author} {\bibinfo {author} {\bibfnamefont {O.}~\bibnamefont
  {{Kop{\'a}{\v{c}}ek}}}, \bibinfo {author} {\bibfnamefont {V.}~\bibnamefont
  {{Karas}}}, \bibinfo {author} {\bibfnamefont {J.}~\bibnamefont
  {{Kov{\'a}{\v{r}}}}},\ and\ \bibinfo {author} {\bibfnamefont
  {Z.}~\bibnamefont {{Stuchl{\'\i}k}}},\ }\href
  {https://doi.org/10.1088/0004-637X/722/2/1240} {\bibfield  {journal}
  {\bibinfo  {journal} {\apj}\ }\textbf {\bibinfo {volume} {722}},\ \bibinfo
  {pages} {1240} (\bibinfo {year} {2010})},\ \Eprint
  {https://arxiv.org/abs/1008.4650} {arXiv:1008.4650 [astro-ph.HE]}
  \BibitemShut {NoStop}%
\bibitem [{\citenamefont {Takens}(1981)}]{Takens:1981emb}%
  \BibitemOpen
  \bibfield  {author} {\bibinfo {author} {\bibfnamefont {F.}~\bibnamefont
  {Takens}},\ }in\ \href@noop {} {\emph {\bibinfo {booktitle} {Dynamical
  Systems and Turbulence, Warwick 1980}}},\ \bibinfo {editor} {edited by\
  \bibinfo {editor} {\bibfnamefont {D.}~\bibnamefont {Rand}}\ and\ \bibinfo
  {editor} {\bibfnamefont {L.-S.}\ \bibnamefont {Young}}}\ (\bibinfo
  {publisher} {Springer Berlin Heidelberg},\ \bibinfo {address} {Berlin,
  Heidelberg},\ \bibinfo {year} {1981})\ pp.\ \bibinfo {pages}
  {366--381}\BibitemShut {NoStop}%
\bibitem [{\citenamefont {Hochreiter}\ and\ \citenamefont
  {Schmidhuber}(1997)}]{Hochreiter:1997lstm}%
  \BibitemOpen
  \bibfield  {author} {\bibinfo {author} {\bibfnamefont {S.}~\bibnamefont
  {Hochreiter}}\ and\ \bibinfo {author} {\bibfnamefont {J.}~\bibnamefont
  {Schmidhuber}},\ }\href {https://doi.org/10.1162/neco.1997.9.8.1735}
  {\bibfield  {journal} {\bibinfo  {journal} {Neural Computation}\ }\textbf
  {\bibinfo {volume} {9}},\ \bibinfo {pages} {1735} (\bibinfo {year} {1997})},\
  \Eprint
  {https://arxiv.org/abs/https://direct.mit.edu/neco/article-pdf/9/8/1735/813796/neco.1997.9.8.1735.pdf}
  {https://direct.mit.edu/neco/article-pdf/9/8/1735/813796/neco.1997.9.8.1735.pdf}
  \BibitemShut {NoStop}%
\bibitem [{\citenamefont {Alfaidi}\ and\ \citenamefont
  {Messenger}(2024)}]{Alfaidi:2024ioo}%
  \BibitemOpen
  \bibfield  {author} {\bibinfo {author} {\bibfnamefont {R.}~\bibnamefont
  {Alfaidi}}\ and\ \bibinfo {author} {\bibfnamefont {C.}~\bibnamefont
  {Messenger}},\ }\href@noop {} {\  (\bibinfo {year} {2024})},\ \Eprint
  {https://arxiv.org/abs/2402.04589} {arXiv:2402.04589 [gr-qc]} \BibitemShut
  {NoStop}%
\bibitem [{\citenamefont {Chirikov}(1971)}]{Chirikov:1971gli}%
  \BibitemOpen
  \bibfield  {author} {\bibinfo {author} {\bibfnamefont {B.~V.}\ \bibnamefont
  {Chirikov}},\ }\href@noop {} {\bibinfo {title} {{Research concerning the
  theory of non-linear resonance and stochasticity}}} (\bibinfo {year}
  {1971})\BibitemShut {NoStop}%
\bibitem [{\citenamefont {Morbidelli}(2002)}]{Morbidelli02}%
  \BibitemOpen
  \bibfield  {author} {\bibinfo {author} {\bibfnamefont {A.}~\bibnamefont
  {Morbidelli}},\ }\href@noop {} {\emph {\bibinfo {title} {Modern celestial
  mechanics: aspects of solar system dynamics}}},\ \bibinfo {edition} {1st}\
  ed.\ (\bibinfo  {publisher} {CRC Press},\ \bibinfo {year} {2002})\BibitemShut
  {NoStop}%
\bibitem [{\citenamefont {Reichl}(2004)}]{reichl2004}%
  \BibitemOpen
  \bibfield  {author} {\bibinfo {author} {\bibfnamefont {L.}~\bibnamefont
  {Reichl}},\ }\href {https://books.google.dk/books?id=hlFkbvZOIAwC} {\emph
  {\bibinfo {title} {The Transition to Chaos: Conservative Classical Systems
  and Quantum Manifestations}}},\ Institute for Nonlinear Science\ (\bibinfo
  {publisher} {Springer},\ \bibinfo {year} {2004})\BibitemShut {NoStop}%
\bibitem [{\citenamefont {Zelenka}\ and\ \citenamefont
  {Lukes-Gerakopoulos}(2017)}]{Zelenka:2017aqn}%
  \BibitemOpen
  \bibfield  {author} {\bibinfo {author} {\bibfnamefont {O.}~\bibnamefont
  {Zelenka}}\ and\ \bibinfo {author} {\bibfnamefont {G.}~\bibnamefont
  {Lukes-Gerakopoulos}},\ }in\ \href@noop {} {\emph {\bibinfo {booktitle}
  {{Workshop on Black Holes and Neutron Stars}}}}\ (\bibinfo {year} {2017})\
  \Eprint {https://arxiv.org/abs/1711.02442} {arXiv:1711.02442 [gr-qc]}
  \BibitemShut {NoStop}%
\bibitem [{\citenamefont {Lukes-Gerakopoulos}\ \emph
  {et~al.}(2008)\citenamefont {Lukes-Gerakopoulos}, \citenamefont {Voglis},\
  and\ \citenamefont {Efthymiopoulos}}]{LukesGerakopoulos2008:4dmap}%
  \BibitemOpen
  \bibfield  {author} {\bibinfo {author} {\bibfnamefont {G.}~\bibnamefont
  {Lukes-Gerakopoulos}}, \bibinfo {author} {\bibfnamefont {N.}~\bibnamefont
  {Voglis}},\ and\ \bibinfo {author} {\bibfnamefont {C.}~\bibnamefont
  {Efthymiopoulos}},\ }\href
  {https://doi.org/https://doi.org/10.1016/j.physa.2007.11.024} {\bibfield
  {journal} {\bibinfo  {journal} {Physica A: Statistical Mechanics and its
  Applications}\ }\textbf {\bibinfo {volume} {387}},\ \bibinfo {pages} {1907}
  (\bibinfo {year} {2008})}\BibitemShut {NoStop}%
\bibitem [{\citenamefont {Froeschlé}\ \emph {et~al.}(2005)\citenamefont
  {Froeschlé}, \citenamefont {Guzzo},\ and\ \citenamefont
  {Lega}}]{Froeschl2005}%
  \BibitemOpen
  \bibfield  {author} {\bibinfo {author} {\bibfnamefont {C.}~\bibnamefont
  {Froeschlé}}, \bibinfo {author} {\bibfnamefont {M.}~\bibnamefont {Guzzo}},\
  and\ \bibinfo {author} {\bibfnamefont {E.}~\bibnamefont {Lega}},\ }\href
  {https://doi.org/10.1007/s10569-004-3834-6} {\bibfield  {journal} {\bibinfo
  {journal} {Celestial Mechanics and Dynamical Astronomy}\ }\textbf {\bibinfo
  {volume} {92}},\ \bibinfo {pages} {243–255} (\bibinfo {year}
  {2005})}\BibitemShut {NoStop}%
\bibitem [{\citenamefont {Kingma}\ and\ \citenamefont {Ba}(2015)}]{Adam}%
  \BibitemOpen
  \bibfield  {author} {\bibinfo {author} {\bibfnamefont {D.~P.}\ \bibnamefont
  {Kingma}}\ and\ \bibinfo {author} {\bibfnamefont {J.}~\bibnamefont {Ba}},\
  }in\ \href {http://arxiv.org/abs/1412.6980} {\emph {\bibinfo {booktitle} {3rd
  International Conference on Learning Representations, {ICLR} 2015, San Diego,
  CA, USA, May 7-9, 2015, Conference Track Proceedings}}},\ \bibinfo {editor}
  {edited by\ \bibinfo {editor} {\bibfnamefont {Y.}~\bibnamefont {Bengio}}\
  and\ \bibinfo {editor} {\bibfnamefont {Y.}~\bibnamefont {LeCun}}}\ (\bibinfo
  {year} {2015})\ \Eprint {https://arxiv.org/abs/1412.6980} {arXiv:1412.6980
  [cs.LG]} \BibitemShut {NoStop}%
\bibitem [{\citenamefont {Suková}\ \emph {et~al.}(2016)\citenamefont
  {Suková}, \citenamefont {Grzedzielski},\ and\ \citenamefont
  {Janiuk}}]{Sukova:2015naa}%
  \BibitemOpen
  \bibfield  {author} {\bibinfo {author} {\bibfnamefont {P.}~\bibnamefont
  {Suková}}, \bibinfo {author} {\bibfnamefont {M.}~\bibnamefont
  {Grzedzielski}},\ and\ \bibinfo {author} {\bibfnamefont {A.}~\bibnamefont
  {Janiuk}},\ }\href {https://doi.org/10.1051/0004-6361/201526692} {\bibfield
  {journal} {\bibinfo  {journal} {A\& A}\ }\textbf {\bibinfo {volume} {586}},\
  \bibinfo {pages} {A143} (\bibinfo {year} {2016})}\BibitemShut {NoStop}%
\bibitem [{\citenamefont {Hegger}\ \emph {et~al.}(1999)\citenamefont {Hegger},
  \citenamefont {Kantz},\ and\ \citenamefont {Schreiber}}]{Hegger:1999tisean}%
  \BibitemOpen
  \bibfield  {author} {\bibinfo {author} {\bibfnamefont {R.}~\bibnamefont
  {Hegger}}, \bibinfo {author} {\bibfnamefont {H.}~\bibnamefont {Kantz}},\ and\
  \bibinfo {author} {\bibfnamefont {T.}~\bibnamefont {Schreiber}},\ }\href
  {https://doi.org/10.1063/1.166424} {\bibfield  {journal} {\bibinfo  {journal}
  {Chaos: An Interdisciplinary Journal of Nonlinear Science}\ }\textbf
  {\bibinfo {volume} {9}},\ \bibinfo {pages} {413} (\bibinfo {year} {1999})},\
  \Eprint
  {https://arxiv.org/abs/https://pubs.aip.org/aip/cha/article-pdf/9/2/413/18301636/413\_1\_online.pdf}
  {https://pubs.aip.org/aip/cha/article-pdf/9/2/413/18301636/413\_1\_online.pdf}
  \BibitemShut {NoStop}%
\bibitem [{\citenamefont {Marwan}(2006)}]{commandline_rps}%
  \BibitemOpen
  \bibfield  {author} {\bibinfo {author} {\bibfnamefont {N.}~\bibnamefont
  {Marwan}},\ }\href@noop {} {\bibinfo {title} {Commandline recurrence plots}}
  (\bibinfo {year} {2006}),\ \bibinfo {note}
  {\url{https://tocsy.pik-potsdam.de/commandline-rp.php}}\BibitemShut {NoStop}%
\bibitem [{\citenamefont {Marwan}(2024)}]{rps_website}%
  \BibitemOpen
  \bibfield  {author} {\bibinfo {author} {\bibfnamefont {N.}~\bibnamefont
  {Marwan}},\ }\href@noop {} {\bibinfo {title} {Recurrence plots and cross
  recurrence plots}} (\bibinfo {year} {2024}),\ \bibinfo {note}
  {\url{http://www.recurrence-plot.tk}}\BibitemShut {NoStop}%
\end{thebibliography}%

\appendix
\section{Codes}

Code used to generate trajectories, train and evaluate the networks has been implemented in Python 3.11.5. The machine learning part has been done using the PyTorch framework version 2.1.0 through Python bindings. A release of the code used in this article is under preparation.

To determine optimal embedding parameters, the programs \verb|mutual| and \verb|false_nearest| from the \verb|TISEAN|~\cite{Hegger:1999tisean} package are used. All recurrence plots and quantifiers are computed using the commandline recurrence plots tool~\cite{commandline_rps, rps_website, Marwan:2007rps}.

\end{document}